\documentclass{edbk}
\usepackage{edbkps}
\usepackage{amsmath,amssymb}
\usepackage{epsfig}
\let\footnote\savefootnote
\let\footnoterule\savefootnoterule 
\normallatexbib

\def\avg#1{\left\langle{#1}\right\rangle}

\def\figh{9.2cm}
\def\ffigh{5.7cm}

\begin{document}

\articletitle{Diamagnetic properties of doped antiferromagnets}

\author{D.\ Veberi\v c$^1$, P.\ Prelov\v sek$^{1,2}$, and H.G.\ Evertz$^3$}
\affil{$^1$J.\ Stefan Institute, SI-1000 Ljubljana, Slovenia\\
$^2$Faculty of Mathematics and Physics, University of Ljubljana, SI-1000
Ljubljana, Slovenia\\
$^3$Institute for Theoretical Physics, Technical University  Graz, AT-8010
Graz, Austria}

\begin{abstract}
Finite-temperature diamagnetic properties of doped antiferromagnets as
modeled by the two-dimensional $t$-$J$ model were investigated by
numerical studies of small model systems. Two numerical methods were
used: the worldline quantum Monte Carlo method with a loop cluster
algorithm (QMC) and the finite-temperature Lanczos method (FTLM),
yielding consistent results. The diamagnetic susceptibility introduced
by coupling of the magnetic field to the orbital current reveals an
anomalous temperature dependence, changing character from diamagnetic
to paramagnetic at intermediate temperatures.
\end{abstract}

\inxx{diamagnetic susceptibility}
The dc orbital susceptibility of the system in the external magnetic
field is
\begin{equation}
\chi_d=-\mu_0\frac{\partial^2 F}{\partial B^2}=
-\frac{\chi_0}{\beta}\left[\frac{1}{Z}
\frac{\partial^2 Z}{\partial\alpha^2}-
\left(\frac{1}{Z}\frac{\partial Z}{\partial\alpha}\right)^2\right],
\label{eq:chidef}
\end{equation}
where $\chi_0=\mu_0e^2a^4/\hbar^2$ and $\alpha=eBa^2/\hbar$. In the
previous studies \cite{vebe} it was realized that results are quite
sensitive to finite-size effects, so we also used the QMC method,
where much larger lattices can be studied.

\begin{figure}
\centering
\mbox{
\epsfig{file=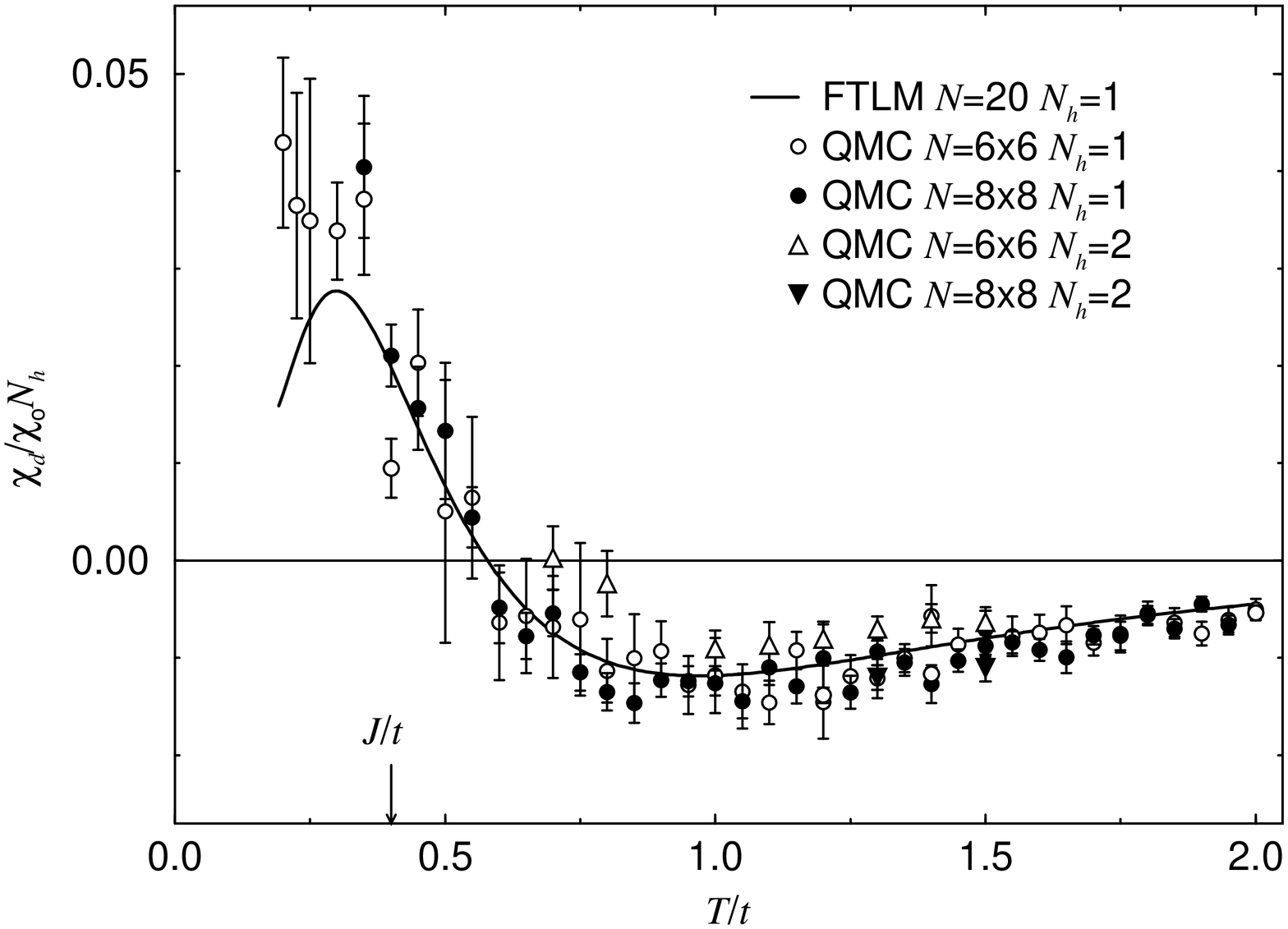,height=\ffigh,angle=-90}
\quad
\epsfig{file=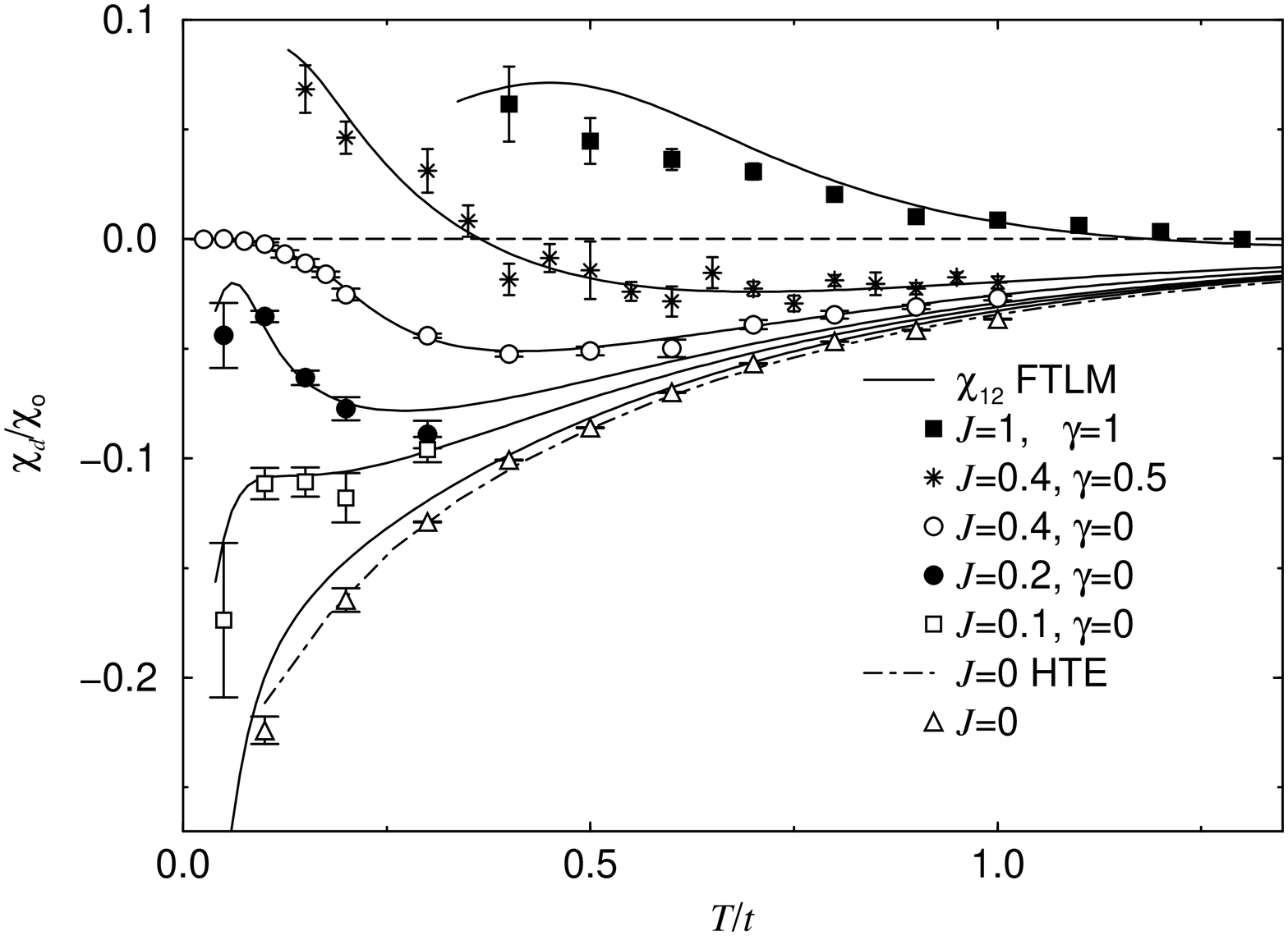,height=\ffigh,angle=-90}
}
\caption[]{Orbital susceptibility $\chi_d$ vs.\ $T$ (left) for one hole
obtained via QMC (dots) and FTLM (line) for $J=0.4t$; different $J$
and $\gamma$ (right). For comparison also results of high-temperature
expansion (HTE) for $J=0$ are shown (dash-dotted line).}
\label{fig:gammaj}
\label{fig:dchi}
\end{figure}

The magnetic field introduced into the $t$-$J$ Hamiltonian via the
Peierls construction, affects only the hopping of the
electrons. Within FTLM the results are obtained by the numerical
derivation with respect to $\alpha$.

Using the standard Trotter-Suzuki decomposition and the {\em woldline}
representation of the QMC for the fermionic models, magnetic field
enters matrix elements concerning the hole hopping. The plaquette
weights along the hole worldline obtain an additional phase
factor. Taking the field derivatives explicitely the expresson for the
orbital susceptibility can be written as
\begin{equation}
\chi_d=-\chi_0\frac{\avg{{\cal S}^2}}{\beta},
\label{eq:chi}
\end{equation}
where $\cal S$ is the projected area of the hole worldline. $\chi_d$
can be thus measured without the presence of a magnetic field. This is
just another consequence of the more general fluctuation--dissipation
theorem. In doped systems we are hindered by the well known
``fermionic sign problem'' of the QMC, not present in the undoped
case. Even though ${\cal S}^2$ is strictly positive, the thermal
average $\avg{{\cal S}^2}$ can become negative because of correlations
between the Monte Carlo sign and the magnitude of the area $\cal
S$. For QMC the sizes of considered systems were $6\times6$ and
$8\times8$.

With FTLM a few mobile holes on a system of tilted squares with up to
20 sites and periodic boundary conditions were considered. It is
nontrivial to incorporate Landau phases corresponding to a homogeneous
$B$, being at the same time compatible with periodic boundary
conditions. This is possible only for quantized magnetic fields.

In Fig.~\ref{fig:dchi}, $\chi_d$ obtained via both methods is
presented. For $T\gg t$, the response is diamagnetic and proportional
to $T^{-3}$ as well as essentially $J$-independent \cite{vebe}. The
most striking effect is that the orbital response below some
temperature $T_p$ turns from diamagnetic to paramagnetic, consistent
with the preliminary results obtained via the FTLM \cite{vebe}. In
order to locate the origin of this phenomenon, results for different
$J$ and anisotropies $\gamma$ are also shown. It appears that $T_p$
scales with $\gamma J$, i.e.\ at $J=0$ the response is clearly
diamagnetic at all $T$, and for $\gamma=0, J>0$ no crossing is
observed with either method.

\begin{figure}
\centering
\epsfig{file=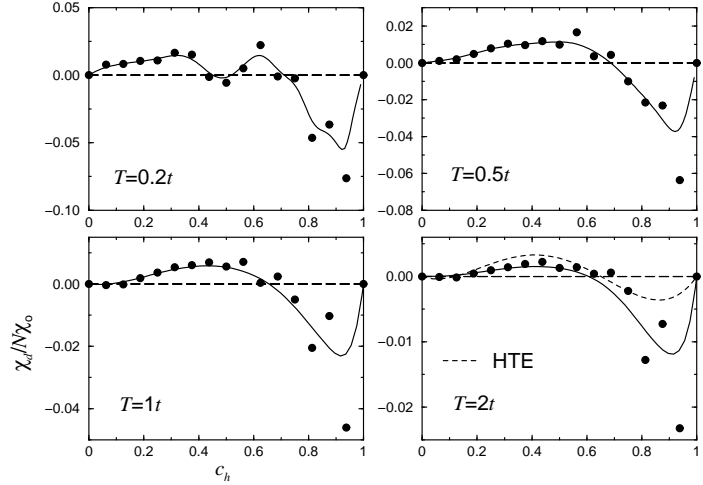,height=\figh,angle=-90}
\caption{$\chi_d$ vs.\ $c_h$ for several $T$ and $J=0.4t$. The last
graph contains also a 4$^{\text{th}}$ order HTE result (dotted).
Canonical (dots) and grand-canonical (line) values for all $c_h$ are
obtained with FTLM on 16 sites.}
\label{fig:chich}
\end{figure}

At lower temperatures $T<T_d\ll T_p$, the diamagnetic behavior is
expected to be restored. This follows from the argument that at
$T\to0$ a hole in an AFM should behave as a quasiparticle with a
finite effective mass, exhibiting a cyclotron motion in $B\neq0$,
leading to $\chi_d(T\to0)\to-\infty$ \cite{vebe}. Numerically it is
easiest to test this conjecture for a single hole and $\gamma=0$. This
is also true for $J=0$.

In Fig.~\ref{fig:chich} also results for $\chi_d$ for finite doping
$c_h>0$ are presented. For nearly empty band $c_h>0.7$ the $\chi_d$ is
diamagnetic and weakly dependent on $T$. In this dilute regime strong
correlations are unimportant, thus Landau diamagnetism is expected. At
moderate temperatures $T>J$ and for an intermediate-doping
$0.2<c_h<0.7$ the $\chi_d$ is dominated by a paramagnetic
response. There is a weak diamagnetism at $c_h<0.2$ and $T>T_p$, while
the paramagnetic regime extends to $c_h=0$ for $T<T_p$. For low
temperatures $T \ll J$ quite pronounced oscillations in $\chi_d(c_h)$
appear and can be partly attributed to finite-system effects.

The explanation can go in the direction proposed by \cite{laug}, that
at low doping $c_h \to 0$ we are dealing with quasiparticles (with a
diamagnetic response), being a bound composite of charge (holon) and
spin (spinon) elementary excitations. The binding appears to be quite
weak and thus easily destroyed by finite $T$ or $c_h$, enabling the
independent and apparently paramagnetic response of constituents.

\end{document}